\documentclass{bioinfo}%
\usepackage{graphicx}
\usepackage{subfigure}
\usepackage{amsmath}
\usepackage{amssymb}
\usepackage{textcomp}
\usepackage{algorithmic}
\usepackage{algorithm}
\usepackage{rotating}
\usepackage{amsfonts}%
\providecommand{\U}[1]{\protect\rule{.1in}{.1in}}
\copyrightyear{2005}
\pubyear{2005}
\begin{document}

\title[SEK]{SEK: Sparsity exploiting $k$-mer-based estimation of bacterial community composition}
\author[Chatterjee et. al.]{Saikat Chatterjee\,$^{1}$\thanks{To whom correspondence should be addressed. Email: sach@kth.se}, 
David Koslicki\,$^{2}$, Siyuan Dong\,$^{3}$, Nicolas Innocenti\,$^{4}$,
Lu Cheng\,$^{5}$, Yueheng Lan\,$^{6}$, Mikko Vehkaper\"a\,$^{1,8}$, Mikael Skoglund\,$^{1}$,  Lars K. Rasmussen\,$^{1}$, Erik Aurell\,$^{4,7}$, Jukka
Corander\,$^{5}$ }

\address{$^{1}$Dept of Communication Theory,  KTH Royal Institute of Technology, Sweden\\ 
$^{2}$Dept of Mathematics, Oregon State University, Corvallis, USA\\
$^{3}$Systems Biology program, KTH Royal Institute of Technology, Sweden and Aalto University, Finland\\ 
$^{4}$Dept of Computational Biology, KTH Royal Institute of Technology, Sweden\\
$^{5}$Dept of Mathematics and Statistics, University of Helsinki, Finland\\ 
$^{6}$Dept of Physics, Tsinghua University, Beijing, China\\ 
$^{7}$Dept of Information and Computer Science, Aalto University, Finland \\
$^{8}$Dept of Signal Processing, Aalto University, Finland\\ } 

\history{Received on XXXXX; revised on XXXXX; accepted on XXXXX}

\editor{Associate Editor: XXXXXXX}

\maketitle

\begin{abstract}
Motivation: Estimation of bacterial community composition from a
high-throughput sequenced sample is an important task in metagenomics
applications. Since the sample sequence data typically harbors reads of
variable lengths and different levels of biological and technical noise,
accurate statistical analysis of such data is challenging. Currently popular
estimation methods are typically very time consuming in a desktop computing environment.

Results: Using sparsity enforcing methods from the general sparse signal processing field 
(such as compressed sensing), we
derive a solution to the community composition estimation problem by a
simultaneous assignment of all sample reads to a pre-processed reference
database. A general statistical model based on kernel density estimation techniques is
introduced for the assignment task and the model solution is obtained using
convex optimization tools. Further, we design a greedy algorithm solution for a fast solution.
Our approach offers a reasonably fast community composition estimation method 
which is shown to be more robust to input data variation than a recently introduced related method.

Availability: A platform-independent Matlab implementation of the method is
freely available at http://www.ee.kth.se/ctsoftware;
source code that does not require access to Matlab is currently being tested
and will be made available later through the above website.

\end{abstract}

\firstpage{1}


\section{Introduction}


%

High-throughput sequencing technologies have recently enabled detection of
bacterial community composition at an unprecedented level of detail.
The high-throughput approach focuses on producing for each sample a large
number of reads covering certain variable part of the 16S rRNA gene, which
enables an identification and comparison of the relative frequencies of
different taxonomic units present across samples. Depending on the
characteristics of the samples, the bacteria involved and the quality of the
acquired sequences, the taxonomic units may correspond to species, genera or
even higher levels of hierarchical classification of the variation existing in
the bacterial kingdom. However, at the same time, the rapidly increasing sizes
of read sets produced per sample in a typical project call for fast inference
methods to assign meaningful labels to the sequence data, a problem which has
attracted considerable attention
\cite{NaiveBayesianClassifier_Wang_2007,meinicke2011mixture,Quikr_Koslicki_2013,Ong_2013}.

Many approaches to the bacterial community composition estimation problem
use 16S rRNA amplicon sequencing where thousands to hundreds of thousands of
moderate length (around 250-500 bp) reads are produced from each sample and
then either clustered or classified to obtain estimates of the prevalence of
any particular taxonomic unit. In the clustering approach the reads are
grouped into taxonomic units by either distance-based or probabilistic methods
\cite{ESPRIT_Tree_Cai_B16SrRNA_2011,Edgar_2010,Cheng_Walker_Corander_2012metagenomic}, 
such that the actual taxonomic
labels are assigned to the clusters afterwards by matching their consensus
sequences to a reference database. Recently, the Bayesian BeBAC method
\cite{Cheng_Walker_Corander_2012metagenomic} was shown to provide high
biological fidelity in clustering. However, this accuracy comes with a
substantial computational cost such that a running time of several days in a
computing-cluster environment may be required for large read sets. In contrast to the
clustering methods, the classification approach is based on using a reference
database directly to assign reads to meaningful units representing biological
variations. Methods for the classification of reads have been based either on
homology using sequence similarity or on genomic signatures in terms of
oligonucleotide composition. Examples of homology-based methods include MEGAN
\cite{Huson_MEGAN_2007,Mitra_rRNA_MEGAN_2011} and phylogenetic analysis
\cite{von_Mering_2007}. A popular approach is the Ribosomal Database
Project's (RDP) classifier which is based on a na\"ive Bayesian classifier (NBC) that 
assigns a label explicitly to each read produced for a particular sample 
\cite{NaiveBayesianClassifier_Wang_2007}. Despite the computational
simplicity of NBC, the RDP classifier may still require several days to
process a data set in a desktop environment. Given this challenge,
considerably faster methods based on different convex optimization strategies
have been recently proposed \cite{meinicke2011mixture,Quikr_Koslicki_2013}. In
particular, sparsity-based techniques, mainly compressive sensing based
algorithms \cite{CS_introduction_Candes_Wakin_2008}, are used for estimation
of bacterial community composition in \cite{amir_CS_2011,Quikr_Koslicki_2013,Zuk_SPIRE_2013}.
However, \cite{amir_CS_2011} used sparsity-promoting algorithms to analyze
mixtures of dye-terminator reads resulting from Sanger sequencing, with the
sparsity assumption that each bacterial community is comprised of a small
subset of known bacterial species, the scope of the work thus being different
from methods intended for high-throughput sequence data. The Quikr method of
\cite{Quikr_Koslicki_2013} uses a $k$-mer-based approach on 16S rRNA sequence
reads and has a considerable similarity to the method (SEK: Sparsity
Exploiting K-mers-based algorithm) introduced here. Explained briefly, the Quikr setup
is based on the following core theoretical formulation: given a reference database
$D=\{d_{1},\ldots,d_{M}\}$ of sequences and a set $S=\{s_{1},\ldots,s_{t}\}$
of sample sequences (the reads to be classified), it is assumed that there
exists a unique $d_{j}$ for each $s_{l}$, such that $s_{l}=d_{j}$. In general,
all reference databases and sample sets consist of sequences with highly
variable lengths. In particular the lengths of reference sequences and samples
reads are often quite different. Violation of the assumption leads to
sensitivity in Quikr performance according to our experiments. 
Another example of fast estimation is called Taxy \cite{meinicke2011mixture} 
that addresses the effect of varying sequence lengths \cite{Wommack_read_lengths_2008}.
Taxy uses a mixture model for the system setting and convex optimization for a solution.
The method referred to as COMPASS \cite{COMPASS_Amir_2013} is another  
convex optimization approach, very similar to the Quikr method, that uses 
large $k$-mers and a divide-and-conquer technique to handle very large 
resulting training matrices. The currently available version of the Matlab-based COMPASS software 
does not allow for training with custom databases, so a direct comparison to SEK is not yet possible.

To enable fast estimation, we adopt an approach where the estimation of the
bacterial community composition is performed jointly, in contrast to the
read-by-read analysis used in the RDP classifier. Our model is based on kernel
density estimators and mixture density models
\cite{Bishop_MachineLearning_Book_2006}, and it leads to solving an
under-determined system of linear equations under a particular sparsity
assumption. In summary, the SEK approach is implemented in three separate
steps: off-line computation of $k$-mers using a reference
database of 16S rRNA genes with known taxonomic classification,
on-line computation of $k$-mers for a given sample, and then final on-line
estimation of the relative frequencies of taxonomic units in the sample by
solving an under-determined system of linear equations.

\section{Methods}

\subsection{General notation and computational resources used}

We denote the non-negative real line by
$\mathbb{R}_{+}$. The $\ell_{p}$ norm is denoted by $\Vert.\Vert_{p}$, and
$\mathbb{E}[.]$ denotes the expectation operator. Transpose of a vector/matrix
is denoted by $(.)^t$. We denote cardinality and complement of a set $\mathcal{S}$
by $|\mathcal{S}|$ and $\overline{\mathcal{S}}$, respectively.
In the computations reported in the remainder of the paper we used standard 
Matlab software with some instances of C code. For experiments on mock community
data, we used a Dell Latitude E6400 laptop computer with a 3 GHz processor and 8 GB memory. 
We also used the {$\mathbf{cvx}$} \cite{Boyd_2004_Book} convex optimization toolbox and the
Matlab function {$\mathbf{lsqnonneg}$()} for a least-squares solution with
non-negativity constraint. For experiments on simulated data, we used standard computers with an 
Intel Xeon x5650 processor and an Intel i7-4930K processor.

\subsection{$k$-mer training matrix from reference data}

\label{subsec:training} 
The training step of SEK consists of converting an
input labeled database of 16S rRNA sequences into a $k$-mer training matrix.
For a fixed $k$, we calculate $k$-mers feature vectors for a window of
fixed length, such that the window is shifted (or slid) by a fixed number of positions
over a database sequence. This procedure captures variability of localized 
$k$-mer statistics along 16S rRNA sequences. Using bp as the
length unit and denoting the length of a reference database sequence $d$ by
$L_{d}$, and further a fixed window length by $L_{w}\leq L_{d}$ and the fixed
position shift by $L_{p}$, the total number of sub-sequences processed to
$k$-mers is close to $\lfloor\frac{L_{d}-L_{w}}{L_{p}}\rfloor$. The choice of $L_{w}$
may be decided by the shortest sample sequence length that is used in the estimation
assuming the reads in a sample set are always shorter than the reference
training sequences. In practice, for example, we used $L_{w}=450$ bp in 
experiments using mock communities data. The choice of $L_{p}$
is decided by the trade-off between computational complexity and
estimation performance.

Given a database of reference training sequences $D=\{d_{1},\ldots,d_{M}\}$ where $d_{m}$ is the
sequence of the $m$th taxonomic unit, each sequence $d_{m}$ is treated
independently. For $d_{m}$, the $k$-mer feature vectors are stored
column-wise in a matrix $\mathbf{X}_{m}\in\mathbb{R}_{+}^{4^{k}\times N_{m}}$,
where $N_{m} \approx \lfloor\frac{L_{d_{m}}-L_{w}}{L_{p}}\rfloor$. From the training
database $D$, we obtain the full training matrix
\[%
\begin{array}
[c]{rl}%
\mathbf{X}= & \left[  \mathbf{X}_{1}\,\mathbf{X}_{2}\,\ldots,\mathbf{X}%
_{M}\right]  \,\,\in\mathbb{R}_{+}^{4^{k}\times N},\\
\equiv & \left[  \mathbf{x}_{1}\,\mathbf{x}_{2}\,\ldots\mathbf{x}_{N}\right]
,
\end{array}
\]
where $\sum_{m=1}^{M}N_{m}=N$, and $\mathbf{x}_{n}\in\mathbb{R}_{+}^{4^{k}\times 1}$ denotes the $n$th $k$-mers
feature vector in the full set of training feature vectors $\mathbf{X}$.

\subsection{SEK model}

For the $m$th taxonomic unit, we have the training set
\[
\mathbf{X}_{m}=\left[  \mathbf{x}_{m1}\,\mathbf{x}_{m2}\,\ldots\mathbf{x}%
_{mN_{m}}\right]  \,\,\in\mathbb{R}_{+}^{{4^k}\times N_{m}},
\]
where we used an alternative indexing to denote the $l$th $k$-mer feature
vector by $\mathbf{x}_{ml}$. Letting $\mathbf{x}$
and $\mathcal{C}_{m}$ denote random $k$-mer feature vectors and $m$th taxonomic unit 
respectively, and using $\mathbf{X}_{m}$, we first model the conditional density
$p(\mathbf{x}|\mathcal{C}_{m})$ corresponding to $m$th unit by a mixture density as
\begin{align}
p(\mathbf{x} | \mathcal{C}_{m}) = \sum_{l=1}^{N_{m}} \alpha_{ml} \,\,
p_{ml}(\mathbf{x} | \mathbf{x}_{ml}, \Theta_{ml}
),
\label{eq:Class_conditional_density}%
\end{align}
where $\alpha_{ml}\geq0$, $\sum_{l=1}^{N_{m}}\alpha_{ml}=1$,
$\mathbf{x}_{ml}$ is assumed to be the mean of distribution $p_{ml}$ 
and $\Theta_{ml}$ denotes the other parameters/properties apart from the mean. In
general, $p_{ml}$ could be chosen according to any convenient parametric
or non-parametric family of distributions. In biological terms, $\alpha_{ml}$ reflects the
amplification of a variable sequence region and how probable that is in a given dataset with a
sufficient level of coverage. The approach of using training data
$\mathbf{x}_{ml}$ as the mean of $p_{ml}$ stems from a standard approach of 
using kernel density estimators (see section 2.5.1 of \cite{Bishop_MachineLearning_Book_2006}). 

Given a test set of $k$-mers (computed from reads), the distribution of the test set is modeled as%
\[
p(\mathbf{x})=\sum_{m=1}^{M}p(\mathcal{C}_{m})\,\,p(\mathbf{x}|\mathcal{C}_{m}),
\]
where we denote probability for taxonomic unit $m$ (or class weight) by $p(\mathcal{C}_{m})$, 
satisfying $\sum_{m=1}^{M}p(\mathcal{C}_{m})=1$. Note that $\{ p(\mathcal{C}_{m}) \}_{m=1}^M$ is 
the composition of taxonomic units. The inference task is
to estimate $p(\mathcal{C}_{m})$ as accurately and fast as possible, for which
a first order moment matching approach is developed. We first evaluate the mean of
$\mathbf{x}$ under $p(\mathbf{x})$ as follows
\[%
\begin{array}
[c]{l}%
\mathbb{E}[\mathbf{x}]\\
=\int\mathbf{x}\,\,p(\mathbf{x})\,d\mathbf{x}\in\mathbb{R}_{+}^{4^k\times1}\\
=\sum_{m=1}^{M}p(\mathcal{C}_{m})\int\mathbf{x}\,\,p(\mathbf{x}|\mathcal{C}_{m})\,d\mathbf{x}\\
=\sum_{m=1}^{M}p(\mathcal{C}_{m})\int\mathbf{x}\,\,\sum_{l=1}^{N_{m}}%
\alpha_{ml}\,\,p_{ml}(\mathbf{x}|\mathbf{x}_{ml},\Theta_{ml})\,d\mathbf{x}\\
=\sum_{m=1}^{M}p(\mathcal{C}_{m})\sum_{l=1}^{N_{m}}\,\,\alpha_{ml}%
\int\mathbf{x}\,\,p_{ml}(\mathbf{x}|\mathbf{x}_{ml},\Theta_{ml})\,d\mathbf{x}%
\\
=\sum_{m=1}^{M}p(\mathcal{C}_{m})\sum_{l=1}^{N_{m}}\alpha_{ml}\,\mathbf{x}%
_{ml}.
\end{array}
\]
Introducing a new indexing $n \triangleq n(m,l)=\sum_{j=1}^{m-1} N_j + l$, we can write
\[
\mathbb{E}[\mathbf{x}]=\sum_{n=1}^{N}\gamma_{n}\,\mathbf{x}_{n}=\mathbf{X}%
\mbox{\boldmath{$\gamma$}},
\]
where 
\begin{align}
\begin{array}{l}
 \mbox{\boldmath{$\gamma$}} = [\gamma_{1}\,\gamma_{2}\,\ldots,\gamma
_{N}]^{T}\in\mathbb{R}_{+}^{N\times1}, \\ 
 \gamma_{n} \triangleq \gamma_{n(m,l)} = p(\mathcal{C}_{m}) \alpha_{ml},
\end{array}
\label{eq:definition_gamma_n}
\end{align}
with the following properties
\[%
\begin{array}
[c]{l}%
\displaystyle\sum_{n(m,1)}^{n(m,N_{m})}
\gamma_{n}=p(\mathcal{C}_{m})\sum_{l=1}^{N_{m}}\alpha_{ml}=p(\mathcal{C}%
_{m}),\\
\sum_{n=1}^{N}\gamma_{n}=\Vert\mbox{\boldmath{$\gamma$}}\Vert_{1}=1.
\end{array}
\]

In our approach we use the sample mean of the test set. The test set consists 
of $k$-mers feature vectors computed from reads. Each read is processed individually 
to generate $k$-mers in the same manner used for the reference data. 
We compute sample mean of the $k$-mer feature vectors for test dataset reads.
Let us denote the sample mean of the test dataset by
$\mbox{\boldmath{$\mu$}}\in\mathbb{R}_{+}^{4^{k}\times1}$, and assume that
the number of reads is reasonably high such that 
$\mbox{\boldmath{$\mu$}}\approx\mathbb{E}[\mathbf{x}]$. Then we can write
\[
\mbox{\boldmath{$\mu$}}\approx\mathbf{X}\mbox{\boldmath{$\gamma$}}.
\]
Considering that model irregularities are absorbed in an additive noise term 
$\mathbf{n}$, we use the following system model
\begin{align}
\mbox{\boldmath{$\mu$}} = \mathbf{X} \mbox{\boldmath{$\gamma$}} + \mathbf{n}
\in\mathbb{R}_{+}^{4^{k} \times1}.
\label{eq:underdetermined_additive_setup_metagenomics}%
\end{align}
Using the sample mean $\mbox{\boldmath{$\mu$}}$ and knowing $\mathbf{X}$, we
estimate $\mbox{\boldmath{$\gamma$}}$ from \eqref{eq:underdetermined_additive_setup_metagenomics} 
as $\hat{\mbox{\boldmath{$\gamma$}}} 
\triangleq [\hat{\gamma}_{1}\,\hat{\gamma}_{2}\,\ldots,\hat{\gamma}_{N}]^{T}
\in\mathbb{R}_{+}^{N\times1}$ followed by estimation of $p(\mathcal{C}_{m})$
as
\[
\hat{p}(\mathcal{C}_{m})=\sum_{n(m,1)}^{n(m,N_{m})}\hat{\gamma}_{n}.
\]
Note that the estimation $\hat{\mbox{\boldmath{$\gamma$}}}\in\mathbb{R}%
_{+}^{N\times1}$ must satisfy the following constraints
\begin{align}
\begin{array}
[c]{l}%
\hat{\mbox{\boldmath{$\gamma$}}}\geq\mathbf{0},\\
\Vert\hat{\mbox{\boldmath{$\gamma$}}}\Vert_{1}=\sum_{n=1}^{N}\hat{\gamma}%
_{n}=\sum_{m=1}^{M}\hat{p}(\mathcal{C}_{m})=1.
\end{array}
\label{eq:solution_constraint_underdetermined_additive_setup_metagenomics}
\end{align}
In \eqref{eq:solution_constraint_underdetermined_additive_setup_metagenomics},
$\hat{\mbox{\boldmath{$\gamma$}}}\geq\mathbf{0}$ means $\forall n$, $\hat{\gamma}_{n} \geq 0$. 
We note that the linear setup
\eqref{eq:underdetermined_additive_setup_metagenomics} is under-determined as
$4^{k}<N$ (in practice $4^{k}\ll N$) and hence, in general, 
solving \eqref{eq:underdetermined_additive_setup_metagenomics} without any constraint 
will lead to infinitely many solutions. 
The constraints \eqref{eq:solution_constraint_underdetermined_additive_setup_metagenomics}
result in a feasible set of solutions that is convex and can be used for finding
a unique and meaningful solution.

We recall that the main interest is to estimate $p(\mathcal{C}_{m})$, which
is achieved in our approach by first estimating $\mbox{\boldmath{$\gamma$}}$
and then $p(\mathcal{C}_{m})$. Hence $\mbox{\boldmath{$\gamma$}}$ represents
an auxiliary variable in our system.

\subsection{Optimization problem and sparsity aspect}
\label{subsec:Optimization_problem}

The solution of \eqref{eq:underdetermined_additive_setup_metagenomics}, denoted by
$\hat{\mbox{\boldmath{$\gamma$}}}$, must satisfy the constraints 
in \eqref{eq:solution_constraint_underdetermined_additive_setup_metagenomics}.
Hence, for SEK, we pose the optimization problem to solve as follows
\begin{equation}
{\rm P}_{\rm sek}^{+,1}: \hspace{0.5cm} \hat{\mbox{\boldmath{$\gamma$}}}=\underset{\mbox{\boldmath{$\gamma$}}}
{\arg\min}\left\Vert \mathbf{\mbox{\boldmath{$\mu$}}
}-\mathbf{X}\mbox{\boldmath{$\gamma$}}\right\Vert _{2},\mbox{\boldmath{$\gamma$}}\geq0,\Vert
\mbox{\boldmath{$\gamma$}}\Vert_{1}=1,
\label{eq:QP_SEK}
\end{equation}
where `$+$' and `$1$' notations in ${\rm P}_{\rm sek}^{+,1}$ refer to the constraints $\hat
{\mbox{\boldmath{$\gamma$}}}\in\mathbb{R}_{+}^{N}$ and $\Vert\hat
{\mbox{\boldmath{$\gamma$}}}\Vert_{1}=1$, respectively. 
The problem ${\rm P}_{\rm sek}^{+,1}$ is a constrained least squares problem and 
a quadratic program (QP) solvable by convex optimization tools, 
such as {$\mathbf{cvx}$} \cite{cvx_toolbox_2013}. In our assumption $4^k < N$, 
and hence the required computation complexity is $\mathcal{O}(N^{3})$ \cite{Boyd_2004_Book}.

The form of ${\rm P}_{\rm sek}^{+,1}$ bears resembance to the widely used LASSO-method
from general sparse signal processing, mainly used for solving 
under-determined problems in compressive sensing
\cite{CS_introduction_Candes_Wakin_2008,Chatterjee_Sundman_Vehkapera_Skoglund_TSP_2012}. 
LASSO deals with the following optimization problem 
(see (1.5) of \cite{Efron_2004_LARS})
\begin{equation}
{\rm LASSO:} \hspace{0.5cm} \hat{\mbox{\boldmath{$\gamma$}}}_{\rm lasso}=\underset{\mbox{\boldmath{$\gamma$}}}
{\arg\min}\left\Vert \mathbf{\mbox{\boldmath{$\mu$}}
}-\mathbf{X}\mbox{\boldmath{$\gamma$}}\right\Vert _{2},\Vert
\mbox{\boldmath{$\gamma$}}\Vert _{1} \leq \tau, \nonumber
\label{eq:QP_SEK}
\end{equation}
where $\tau \in \mathbb{R}_{+}$ is a user choice that decides the level of sparsity in 
$\hat{\mbox{\boldmath{$\gamma$}}}_{\rm lasso}$; for example $\tau=1$ will lead
to a certain level of sparsity. A decreasing $\tau$ leads to an increasing
level of sparsity in LASSO solution. LASSO is often presented in an unconstrained
Lagrangian form that minimizes $\{ \left\Vert \mathbf{\mbox{\boldmath{$\mu$}}
}-\mathbf{X}\mbox{\boldmath{$\gamma$}}\right\Vert _{2}^{2} + \lambda \Vert
\mbox{\boldmath{$\gamma$}}\Vert _{1}\}$, where $\lambda$ decides the level of sparsity.
${\rm P}_{\rm sek}^{+,1}$ is not theoretically 
bound to provide a sparse solution with a similar level of sparsity achieved by LASSO when a small $\tau < 1$ is used.

For the community composition estimation problem, the auxiliary variable 
$\mbox{\boldmath{$\gamma$}}$ defined in \eqref{eq:definition_gamma_n}
is inherently sparse. Two particularly natural motivations concerning the sparsity can
be brought forward. Firstly, consider the conditional densities for taxonomic
units as shown in \eqref{eq:Class_conditional_density}. Regarding the
conditional density model for a single unit, a natural hypothesis for the
generating model is that the conditional densities for several other units
will induce only few feature vectors, and hence $\alpha_{ml}$ will be 
negligible or effectively zero for certain patterns in the
feature space, leading to sparsity in the auxiliary variable
$\mbox{\boldmath{$\gamma$}}$ (unstructured sparsity
in $\mbox{\boldmath{$\gamma$}}$). Secondly, in most samples only a small 
fraction of the possible taxonomic units is expected to be present, 
and consequently, many $p(\mathcal{C}_{m})$ will turn out to be zero, which
again corresponds to sparsity in $\mbox{\boldmath{$\gamma$}}$ (structured block-wise sparsity in $\mbox{\boldmath{$\gamma$}}$) 
\cite{Stojnic_2010_Block_Sparse_JSTSP}.
In practice, for a highly under-determined system 
\eqref{eq:underdetermined_additive_setup_metagenomics} in the community composition 
estimation problem with the fact that $\mbox{\boldmath{$\gamma$}}$ is inherently sparse, 
the solution of ${\rm P}_{\rm sek}^{+,1}$ turns out to be 
effectively sparse due to the constraint $\Vert\mbox{\boldmath{$\gamma$}}\Vert_{1}=1$.

\subsection{A greedy estimation algorithm}

For SEK we solve ${\rm P}_{\rm sek}^{+,1}$ using convex optimization tools requiring 
computational complexity $\mathcal{O}(N^{3})$. To reduce the complexity
without a significant loss in estimation performance  we also develop a new
greedy algorithm based on orthogonal matching pursuit (OMP)
\cite{Tropp_2007_OMP}, for a short discussion of OMP
with pseudo-code, see also
\cite{Chatterjee_Sundman_Vehkapera_Skoglund_TSP_2012}. 
In the recent literature several
algorithms have been designed by extending OMP, such as, for example, 
the backtracking based OMP \cite{Makur_BAOMP_2011}, and, by a subset
of the current authors, the look-ahead OMP~\cite{Chatterjee_Sundman_Skoglund_2011_ICASSP_1}.
Since the standard OMP
uses a least-squares approach and does not provide solutions satisfying
constraints in
\eqref{eq:solution_constraint_underdetermined_additive_setup_metagenomics}, it
is necessary to design a new greedy algorithm for the problem addressed here.

The new algorithm introduced here is referred to as ${\rm OMP}_{\rm sek}^{+,1}$, 
and its pseudo-code is shown in Algorithm \ref{alg:OMP_plus_1}. 
In the stopping condition (step~\ref{step:Stop_Condition_OMP_plus_1}), the parameter $\nu$
is a positive real number that is used as a threshold and the parameter $I$ 
is a positive integer that is used to limit the number of iterations. 
The choice of $\nu$ and $I$ is ad-hoc, depending mainly on user experience.
\begin{algorithm}[ht!]
\caption{: ${\rm OMP}_{\rm sek}^{+,1}$ \label{alg:OMP_plus_1} }
\mbox{\textit{Input:}}
\begin{algorithmic}[1]
\STATE $\mathbf{X}$, $\mbox{\boldmath{$\mu$}}$, $\nu$, $I$;
\end{algorithmic}
\mbox{\textit{Initialization:}}
\begin{algorithmic}[1]
\STATE $\mathbf{r}_{0} \leftarrow \mbox{\boldmath{$\mu$}}$, 
$\mathcal{S}_{0} \leftarrow \emptyset$, $i \leftarrow 0$;
\end{algorithmic}
\mbox{\textit{Iterations:}}
\begin{algorithmic}[1]
\REPEAT
\STATE $i \leftarrow i+1$; \hfill (Iteration counter)
\STATE $\tau_{i} \leftarrow $ index of the highest positive element of $\mathbf{X}^{t} \mathbf{r}_{i-1}$;
\label{step:atom_index_OMP_plus_1}
\STATE $\mathcal{S}_{i} \leftarrow \mathcal{S}_{i-1} \cup \tau_{i}$; \hfill ($|\mathcal{S}_{i}|=i$)
\label{step:Intermediate_Support_OMP_plus_1}
\STATE $\tilde{\mbox{\boldmath{$\gamma$}}}_{i} \leftarrow
\underset{{\mbox{\boldmath{$\beta$}}}_{i}} {\arg\min} 
\hspace{2pt} \| \mbox{\boldmath{$\mu$}} -  \mathbf{X}_{\mathcal{S}_{i}} {\mbox{\boldmath{$\beta$}}}_{i} \|_{2},
\hspace{2pt} {\mbox{\boldmath{$\beta$}}}_{i} \geq \mathbf{0}$; 
\hfill ($\mathbf{X}_{\mathcal{S}_{i}}\in\mathbb{R}_{+}^{4^{k} \times i}$)
\label{step:Proj_Residue_OMP_plus_1}
\STATE $\mathbf{r}_{i} \leftarrow
\mbox{\boldmath{$\mu$}} -  \mathbf{X}_{\mathcal{S}_{i}} \tilde{\mbox{\boldmath{$\gamma$}}}_{i}$;
\hfill (Residual)
\label{step:Proj_Residue_OMP_plus_2}
\UNTIL $( ( | \| \tilde{\mbox{\boldmath{$\gamma$}}} \|_1 - 1| \leq \nu) \,\, \mathrm{or}\,\, (i \geq I) )$
\label{step:Stop_Condition_OMP_plus_1}
\end{algorithmic}
\mbox{\textit{Output:}}
\begin{algorithmic}[1]
\STATE $\hat{\mbox{\boldmath{$\gamma$}}} \in \mathbb{R}_{+}^{N}$, satisfying
$\hat{\mbox{\boldmath{$\gamma$}}}_{\mathcal{S}_{i}} = \tilde{\mbox{\boldmath{$\gamma$}}}_{i}$
and $\hat{\mbox{\boldmath{$\gamma$}}}_{\overline{\mathcal{S}}_{i}} = \mathbf{0}$.
\STATE $\hat{\mbox{\boldmath{$\gamma$}}} \leftarrow \frac{\hat{\mbox{\boldmath{$\gamma$}}}}{\| \hat{\mbox{\boldmath{$\gamma$}}} \|_{1}}$
\hfill (Enforcing $\| \hat{\mbox{\boldmath{$\gamma$}}} \|_{1} = 1$)
\label{eq:rescaling_l1_norm_1}
\end{algorithmic}
\end{algorithm}

Compared to the standard OMP, the new aspects in ${\rm OMP}_{\rm sek}^{+,1}$ are as follows:
\begin{itemize}
\item In step \ref{step:atom_index_OMP_plus_1} of $\mbox{\textit{Iterations}}$, we only 
search within positive inner product coefficients.
\item In step \ref{step:Proj_Residue_OMP_plus_1} of $\mbox{\textit{Iterations}}$, 
a least-squares solution $\tilde{\mbox{\boldmath{$\gamma$}}}_{i}$ 
with non-negativity constraint is found for $i$th iteration via the use of intermediate variable
${\mbox{\boldmath{$\beta$}}}_{i} \in\mathbb{R}_{+}^{i \times 1}$. 
In this step, $\mathbf{X}_{\mathcal{S}_{i}}$ is the sub-matrix formed by columns of $\mathbf{X}$
indexed in $\mathcal{S}_{i}$. The concerned optimization problem is convex.
We used the Matlab function {$\mathbf{lsqnonneg}$()} for this purpose.
\item In step \ref{step:Proj_Residue_OMP_plus_2} of $\mbox{\textit{Iterations}}$,
we find the least squares residual $\mathbf{r}_{i}$.
\item In step \ref{step:Stop_Condition_OMP_plus_1} of $\mbox{\textit{Iterations}}$, 
the stopping condition provides for a solution that has an $\ell_{1}$ norm close to one, with an
error decided by the threshold $\nu$. An unconditional stopping condition is provided
by the maximum number of iterations $I$. 
\item In step \ref{eq:rescaling_l1_norm_1} of $\mbox{\textit{Output}}$, the $\ell_{1}$ norm 
of the solution is set to one by a rescaling.
\end{itemize}

The computational complexity of the ${\rm OMP}_{\rm sek}^{+,1}$ algorithm is as follows.
The main cost is incurred at step \ref{step:Proj_Residue_OMP_plus_1} where we
need to solve a linearly constrained quadratic program using convex
optimization tools; here we assume that the costs of the other steps are
negligible. In the $i$th iteration $\mathbf{X}_{\mathcal{S}_{i}}\in\mathbb{R}_{+}^{4^{k}\times i}$
and $i \ll 4^{k}$, and the complexity required to solve step
\ref{step:Proj_Residue_OMP_plus_1} is $\mathcal{O}(4^{k} i^{2})$ \cite{Boyd_2004_Book}. As we have a
stopping condition $i\leq I$, the total complexity of the ${\rm OMP}_{\rm sek}^{+,1}$ algorithm is
within $\mathcal{O}(I\times 4^{k} I^{2})=\mathcal{O}(4^{k} I^{3})$. We know that optimal
solution of ${\rm P}_{\rm sek}^{+,1}$ using convex
optimization tools requires a complexity of $\mathcal{O}(N^{3})$. For a setup with
$I<4^{k}\ll N$, we can have $\mathcal{O}(4^{k} I^{3})\ll\mathcal{O}(N^{3})$, and
hence the ${\rm OMP}_{\rm sek}^{+,1}$ algorithm is typically much more efficient than 
using convex optimization tools directly in a high-dimensional setting.
It is clear that the ${\rm OMP}_{\rm sek}^{+,1}$ algorithm is not allowed to iterate beyond the
limit of $I$; in practice this works as a forced convergence. 
For both ${\rm OMP}_{\rm sek}^{+,1}$ and ${\rm P}_{\rm sek}^{+,1}$, we do not have 
a theoretical proof on robust reconstruction of solutions. Further a natural
question remains on how to set the input parameters $\nu$ and $I$. 
The choice of parameters is discussed later in section~\ref{subsec:Remarks_on_parameter_choice}.

\subsection{Overall system flow-chart}
\begin{figure}[ptb]
\centering
\includegraphics[width=3in,height=2.9in]{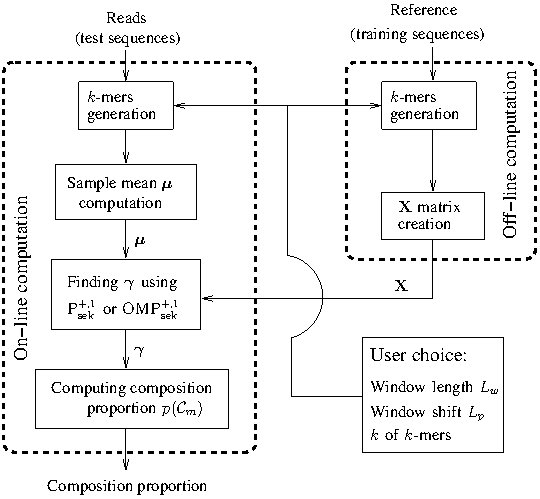}
\caption{A flow-chart of full SEK system.}%
\label{fig:FlowChart}%
\end{figure}

Finally we depict the full SEK system by using a flow-chart shown in Figure~\ref{fig:FlowChart}.
The flow-chart shows main parts of the overall system, and associated off-line and on-line computations.

\subsection{Mock communities data}

For our experiments on real biological data, we used the mock microbial
communities database developed in \cite{Haas_2011}. The database is called
even composition Mock Communities (eMC) for chimeric sequence detection where
the involved bacterial species are known in advance. Three regions (V1-V3, V3-V5, V6-V9)
of the 16S rRNA gene of the composition eMC were sequenced using 454
sequencing technology in four different sequencing centers. In our experiments
we focused on the V3-V5 region datasets, since these have been earlier used
for evaluation of the BeBAC method (see Experiment 2 of
\cite{Cheng_Walker_Corander_2012metagenomic}). 

\subsubsection{Test dataset (Reads):}
\label{subsec:Mock_Test_dataset}

Our basic test dataset used under a variety of different \textit{in silico}
experimental conditions is the one used in Experiment 2 of BeBAC
\cite{Cheng_Walker_Corander_2012metagenomic}. The test dataset consists of
91240 short length reads from 21 different species. The length of reads has a range  
between 450-550 bp and the bacterial community composition is known at the species 
level, by the following computation performed in \cite{Cheng_Walker_Corander_2012metagenomic}. 
Each individual sequence of the 91240 read sequences was aligned 
(local alignment) to all the reference sequences of reference database 
$D^{\rm mock}_{\rm known}$ described in the section~\ref{subsec:Mock_Training_dataset} 
and then each read sequence is labelled by the species of the highest scoring reference sequence, 
followed by computation of the community composition referred to as ground truth.

\subsubsection{Training datasets (Reference):}
\label{subsec:Mock_Training_dataset}

We used two different databases (known and mixed) generated from 
the mock microbial community database
\cite{Haas_2011}. The first database is denoted by $D^{\rm mock}_{\rm known}$ and it consists
of the same $M=21$ species present among the reads described 
in section~\ref{subsec:Mock_Test_dataset}.
The details of the
$D^{\rm mock}_{\rm known}$ database can be found in Experiment 2 of
\cite{Cheng_Walker_Corander_2012metagenomic}. The database consists of 113
reference sequences for a total of 21 bacterial species, such that each
reference sequence represents a distinct 16S rRNA gene. Thus there is a
varying number of reference sequences for each of the considered species. Each
reference sequence has an approximate length of 1500 bp, and for each species, the
corresponding reference sequences are concatenated to a single sequence. The
final reference database $D^{\rm mock}_{\rm known}$ then consists of 21 sequences 
where each sequence has an approximate length 5000 bp.

To evaluate influence of new species in reference data on the performance of SEK, we created
new databases denoted by $D^{\rm mock}_{\rm mixed}(E)$. Here $E$ represents the number of
additional species included to a partial database created from $D^{\rm mock}_{\rm known}$, by
downloading additional reference data from the RDP database.
Each partial database includes only one randomly chosen reference sequence for each
species in $D^{\rm mock}_{\rm known}$ and hence consists of 21 reference sequences 
of approximate length 1500 bp. 
For example, with $E=10$, 10 additional species were included in the reference
database and consequently $D^{\rm mock}_{\rm mixed}(10)$ contains 16S rRNA sequences of
$M=21+10=31$ species. Several instances of 
$D^{\rm mock}_{\rm mixed}(E)$ were made for each fixed value of $E$ by 
choosing a varying set of additional species and we also increased $E$ 
from zero to 100 in steps of 10.
Note that, in $D^{\rm mock}_{\rm mixed}(E)$, the inclusion of only single reference sequence results in reduction of biological variability for 
each of the original 21 species compared to $D^{\rm mock}_{\rm known}$.

\subsection{Simulated data}
\label{section:SimulatedData}

To evaluate how SEK performs for much larger data than 
the mock communities data, we performed experiments for simulated data
described below.

\subsubsection{Test datasets (Reads):}
Two sets of simulated data were used to test the performance of the SEK method. 
First, the 216 different simulated datasets produced in \cite{Quikr_Koslicki_2013} 
were used for a direct comparison to the Quikr method and 
the Ribosomal Database Project's (RDP) Na\"{i}ve Bayesian Classifier (NBC). 
See \cite[\S 2.5]{Quikr_Koslicki_2013} for the design of these simulations. 

The second set of simulated data consists of 486 different pyrosequencing 
datasets constituting over 179M reads generated using the shotgun/amplicon 
read simulator Grinder \citep{Angly2012}. Read-length distributions were 
set to be one of the following: fixed at 100bp, normally distributed 
at $450{\rm bp}\pm 50{\rm bp}$, or normally distributed 
at $800{\rm bp}\pm 100{\rm bp}$. Read depth was fixed to be one of 
10K, 100K, or 1M total reads. Primers were chosen to target either 
only the V1-V3 regions, only the V6-V9 regions, or else the multiple 
variable regions V1-V9. Three different diversity values were chosen 
($10,\ 100$, and $500$) at the species level, and abundance was 
modeled by one of the following three distributions: uniform, linear, 
or power-law with parameter 0.705. Homopolymer errors were modeled 
using Balzer's model \cite{Balzer2010}, and chimera percentages were 
set to either $5\%$ or $35\%$. Since only amplicon sequencing is considered, 
copy bias was employed, but not length bias.

\subsubsection{Training datasets (Reference):}
To analyze the simulated data, two different training matrices were used 
corresponding to the databases $D_{\rm small}$ and $D_{\rm large}$ from \cite{Quikr_Koslicki_2013}. 
The database $D_{\rm small}$ is identical to RDP's NBC training set 7 and 
consists of 10,046 sequences covering 1,813 genera. 
Database $D_{\rm large}$ consists of a 275,727 sequence subset of RDP's 
16S rRNA database covering 2,226 genera. Taxonomic information was obtained from NCBI.

\section{Results}

\subsection{Performance measure and competing methods}

As a quantitative performance measure, we use variational distance (VD) to
compare between known proportions of taxonomic units $\mathbf{p}%
=[p(\mathcal{C}_{1}),\,p(\mathcal{C}_{2}),\ldots,p(\mathcal{C}_{K})]^{T}$ and
the estimated proportions $\hat{\mathbf{p}}=[\hat{p}(\mathcal{C}_{1}),\,\hat
{p}(\mathcal{C}_{2}),\ldots,\hat{p}(\mathcal{C}_{K})]^{T}$. The VD is defined
as
\[
\mathrm{{VD}=0.5\times\Vert\mathbf{p}-\hat{\mathbf{p}}\Vert_{1}\in
\lbrack0,1].}%
\]
A low VD indicates more satisfactory performance.

We compare performances between SEK, Quikr, Taxy and RDP's NBC, for real biological data
(mock communities data) and large size simulated data.

\subsection{Results for Mock Communities data}

Using mock communities data, we carried out experiments where
the community composition problem is addressed at the species level. Here
we investigated how the SEK performs for real biological data, also
vis-a-vis relevant competing methods.

\subsubsection{$k$-mers from test dataset:}
In the test dataset, described in section \ref{subsec:Mock_Test_dataset}, 
the shortest read is of length 450 bp. We used a window length $L_{w}=450$ bp and
refrained from the sliding-the-window approach in the generation of $k$-mers
feature vectors. For $k=4$ and $k=6$, the $k$-mers generation took 21 minutes
and 48 minutes, respectively.

\subsubsection{Results using small training dataset:}
\begin{figure*}[ptb]
\centering
\includegraphics[width=6.5in,height=4.8in]{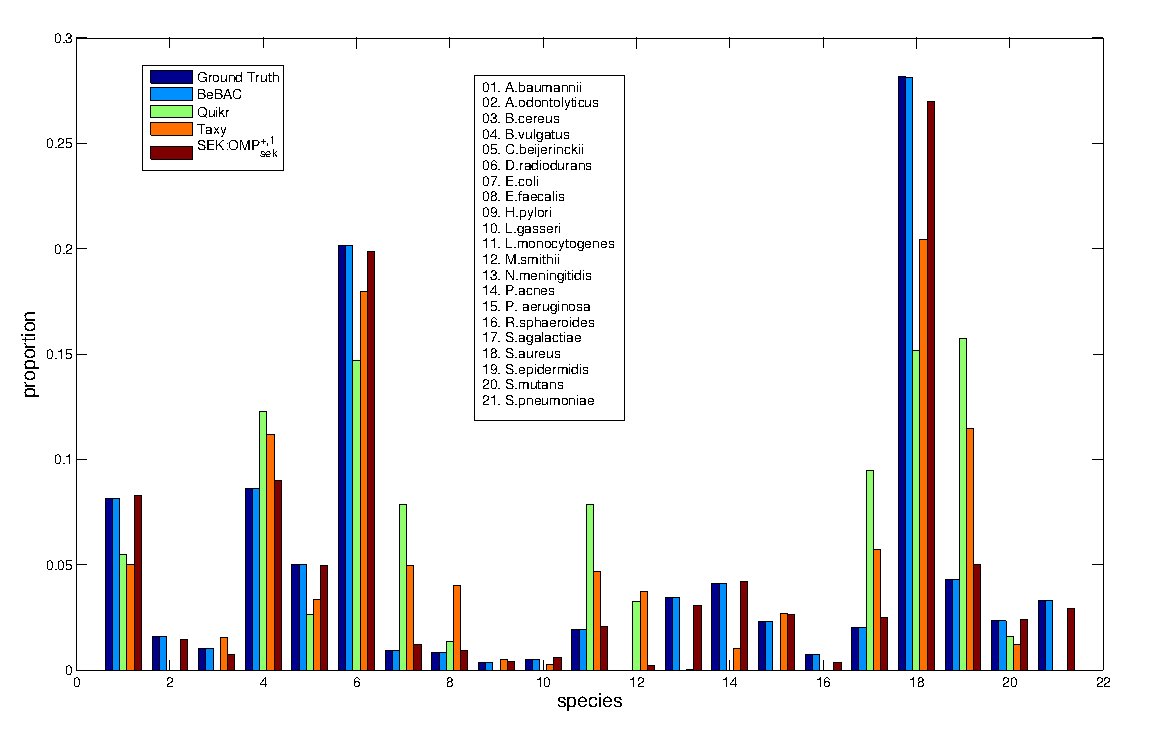} \vspace{-2mm}
\vspace{-3mm}\caption{For mock communities data: Performance of ${\rm OMP}_{\rm sek}^{+,1}$ using reference training database
$D^{\rm mock}_{\rm known}$. Community composition problem is addressed at the species level.
The ${\rm OMP}_{\rm sek}^{+,1}$ performance is shown against the ground truth and
performances of BeBAC, Quikr and Taxy. The ${\rm OMP}_{\rm sek}^{+,1}$ provides better 
match to the ground truth than the competing faster methods Quikr and Taxy.
The corresponding variational distance (VD) performances of BeBAC, ${\rm OMP}_{\rm sek}^{+,1}$,
Taxy and Quikr are 0.0038, 0.0305, 0.2817 and 0.4044, respectively. }%
\label{fig:SEK_comparison}%
\end{figure*}

In this experiment, we used SEK for estimation of the proportions of species in
the test set described in Section~\ref{subsec:Mock_Test_dataset}. Here we used the
smaller training reference set $D^{\rm mock}_{\rm known}$ described in Section~\ref{subsec:training}.
The experimental setup is the same as shown in Experiment 2 of BeBAC
\cite{Cheng_Walker_Corander_2012metagenomic}. Therefore we can directly
compare with the BeBAC results reported in \cite{Cheng_Walker_Corander_2012metagenomic}. 
SEK estimates were based on $4$-mers computed
with the setup $L_{w}=450$ bp and $L_{p}=1$ bp. The choice of $L_{p}=1$ bp 
corresponds to the best case of generating training matrix $\mathbf{X}$, 
with the highest amount of variability in reference $k$-mers. 
Using $D^{\rm mock}_{\rm known}$, the $k$-mers training matrix 
$\mathbf{X}$ has the dimension $4^4\times121412$. For the use of 
SEK in such a high dimension, the QP ${\rm P}_{\rm sek}^{+,1}$ using $\mathbf{cvx}$ 
suffered of numerical instability, but ${\rm OMP}_{\rm sek}^{+,1}$ provided results 
in 3.17 seconds, leading to a VD
= 0.0305. For ${\rm OMP}_{\rm sek}^{+,1}$, $\nu$ and $I$ in algorithm 1 were set to $10^{-5}$ 
and $100$ respectively; the values of these two parameters 
remained unchanged for other experiments
on mock communities data presented later. The performance of SEK using ${\rm OMP}_{\rm sek}^{+,1}$ is shown in
Figure~\ref{fig:SEK_comparison}, and compared against the estimates from
BeBAC, Quikr and Taxy. The Quikr method used $6$-mers and provided a VD =
0.4044, whereas the Taxy method used $7$-mers and provided a VD = 0.2817. The
use of $k=6$ and $k=7$ for Quikr and Taxy, respectively, is chosen according
to the experiments described in \cite{Quikr_Koslicki_2013} and
\cite{meinicke2011mixture}. Here Quikr is found to provide the least
satisfactory performance in terms of VD. BeBAC results are highly accurate
with VD = 0.0038, but come with the requirement of a computation time in the
order of more than thirty hours. On the other hand ${\rm OMP}_{\rm sek}^{+,1}$
had a total online computation time around 21 minutes that is mainly dominated by 
$k$-mers computation from sample reads for evaluating $\mbox{\boldmath{$\mu$}}$; 
given pre-computed $\mathbf{X}$ and $\mbox{\boldmath{$\mu$}}$, the central 
inferenece (or estimation) task of ${\rm OMP}_{\rm sek}^{+,1}$ took only 3.17 seconds.
Considering that Quikr and Taxy also have similar online complexity requirement to 
compute $k$-mers from sample reads, ${\rm OMP}_{\rm sek}^{+,1}$ can be concluded to 
provide a good trade-off between performance and computational demands.

\subsubsection{Results for dimension reduction by higher shifts:}

The $L_{p}=1$ bp leads to a relatively high dimension of $\mathbf{X}$, which
is directly related to an increase in computational complexity. Clearly, the
$L_{p}=1$ bp shift produces highly correlated columns in $\mathbf{X}$ and
consequently it might be sufficient to utilize $k$-mers feature
vectors with a higher shift without a considerable loss in variability
information. To investigate this, we performed an experiment with a gradual
increase in $L_{p}$. We found that selecting $L_{p}=15$ bp results in an input
$\mathbf{X}\in\mathbb{R}_{+}^{4^4\times8052}$ which the $\mathbf{cvx}$ based
${\rm P}_{\rm sek}^{+,1}$ was able to process successfully. At $L_{p}=15$ bp, 
the ${\rm P}_{\rm sek}^{+,1}$ provided a
performance of VD = 0.033260, while the execution time was 25.25 seconds. The
${\rm OMP}_{\rm sek}^{+,1}$ took 1.86 seconds and provided VD = 0.03355l, indicating almost
no performance loss compared to the optimal ${\rm P}_{\rm sek}^{+,1}$. A shift $L_{p}>25$ did result
in a performance drop, for example, $L_{p}=30,50,100$ resulted in
VD values 0.0527, 0.0879, 0.1197, respectively. Therefore, shifts around
$L_{p}=15$ bp appear to be sufficient to reduce the dimension of $\mathbf{X}%
$, while maintaining sufficient biological variability. Hence the
next experiment (in section~\ref{sec:Results_for_mixed_training_dataset}) was conducted using $L_{p}=15$ bp.

\subsubsection{Results for mixed training dataset:}
\label{sec:Results_for_mixed_training_dataset}

\begin{figure*}[ptb]
\centering
\includegraphics[width=6in,height=2.5in]{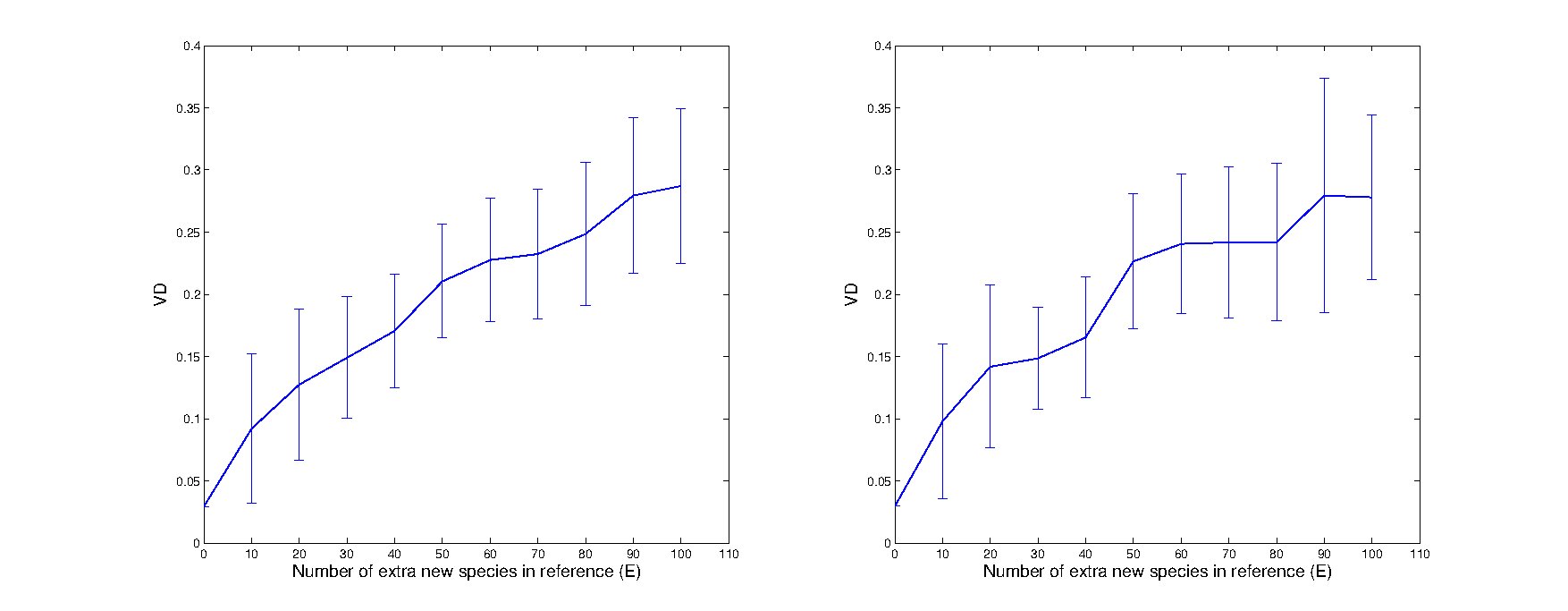}
\caption{For mock communities data: Variational distance (VD) performance of SEK against 
increasing reference database $D^{\rm mock}_{\rm mixed}(E)$, where $E=0,10,20,\ldots,100$. 
The left figure is for ${\rm P}_{\rm sek}^{+,1}$ and the right figure is for ${\rm OMP}_{\rm sek}^{+,1}$. The 
results show that both SEK implementations are subjected to a gradual decrease in performance 
with the increase in the number of additional species.}%
\label{fig:VD_for_increasing_ordered_reference}%
\end{figure*}

In this experiment, we investigated how the performance of SEK varies with an
increase in the number of additional species in the reference training database
which are not present in the sample test data. 
We used reference training datasets $D^{\rm mock}_{\rm mixed}(E)$ 
described in Section~\ref{subsec:training}, where $E=0,10,20,\ldots,100$. 
For each non-zero $E$, we created 10 reference datasets to evaluate variability 
of the performance. The performance with one-sigma error bars is shown 
in Figure~\ref{fig:VD_for_increasing_ordered_reference}. The trend in 
the performance curves confirms that the SEK is subjected to gradual
decrease in performance with the increase in the number of additional species; the trend holds 
for both ${\rm P}_{\rm sek}^{+,1}$ and ${\rm OMP}_{\rm sek}^{+,1}$. 
Also, being optimal the performance of QP ${\rm P}_{\rm sek}^{+,1}$ is found 
to be more consistent than the greedy ${\rm OMP}_{\rm sek}^{+,1}$.

\subsection{Results for Simulated Data}

The simulated data experiments deal with community composition problem at 
different taxonomic ranks and also with very large size of $\mathbf{X}$ in 
\eqref{eq:underdetermined_additive_setup_metagenomics}. 
Due to the massive size of $\mathbf{X}$, a direct application of QP ${\rm P}_{\rm sek}^{+,1}$ 
is not feasible, and hence we used only ${\rm OMP}_{\rm sek}^{+,1}$.
For all results described, $\nu$ and $I$ in algorithm 1 were set to $10^{-5}$ 
and $409$ respectively.

\subsubsection{Training matrix construction:}
\label{sec:SimulatedData_Training_matrix_construction}
In forming the training matrix for $D_{\rm small}$, the $k$-mer size was fixed at $k=6$, 
and the window length and position shifts were set to $L_w=400$ and $L_p=100$ respectively. 
This resulted in a matrix $\mathbf{X}$ with dimensions $4^6 \times 109,773$. 
For the database $D_{\rm large}$, a training matrix $\mathbf{X}$ with dimensions 
$4^6\times 500,734$ was formed by fixing $k=6, L_w=400$, and $L_p=400$. 
Calculating the matrices took $\sim 2.5$ and $\sim 11$ 
minutes respectively using an Intel i7-4930K processor and a custom C program. 
Slightly varying $L_p$ and $L_w$ did not significantly change 
the results contained in sections \ref{section:SimulatedDataResults_1} and 
\ref{section:SimulatedDataResults_2} below, 
but generally decreasing $L_p$ and $L_w$ results in lower reconstruction error 
at the expense of increased execution time and memory usage. 
The values of $L_p$ and $L_w$ were chosen to provide an acceptable 
balance between execution time, memory usage, and reconstruction error.

\subsubsection{Results for first set of simulated data:}
\label{section:SimulatedDataResults_1}

\begin{figure}[!t]
\centering
(a){\includegraphics[width=\columnwidth]{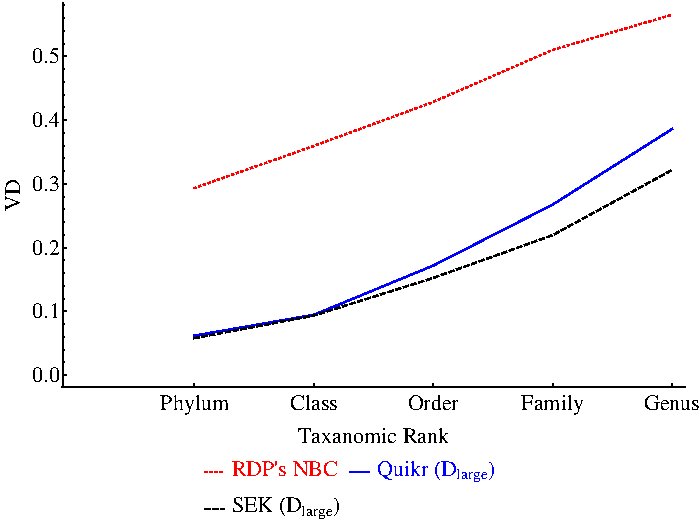}}
(b){\includegraphics[width=\columnwidth]{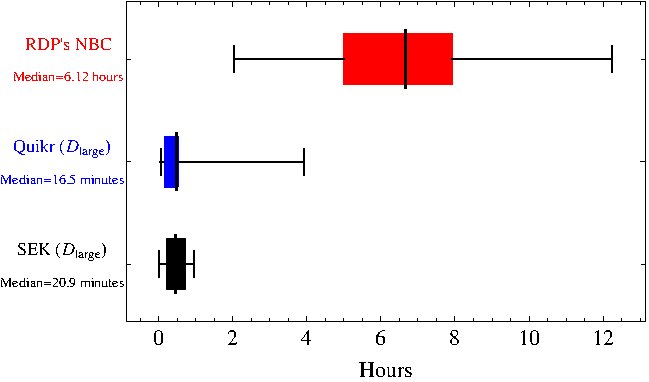}}
\caption{For simulated data: Comparison of SEK (${\rm OMP}_{\rm sek}^{+,1}$) to Quikr 
and RDP's NBC on the first set of simulated data. 
Throughout, RDP's NBC version 10.28 with training set 7 was utilized. 
(a) Variational distance error averaged over all 216 simulated datasets versus 
taxonomic rank for RDP's NBC, with SEK and Quikr trained using $D_{\rm large}$. 
(b) Algorithm execution time for RDP's NBC, with SEK and Quikr trained using 
$D_{\rm large}$. Whiskers denote range of the data, vertical black bars designate 
the median, and the boxes demarcate quantiles.}\label{fig:OldGrinderResults}		
\end{figure}

For test data, $k$-mers were computed in the same manner as described in 
section~\ref{sec:SimulatedData_Training_matrix_construction}.
On average, 4.0 seconds were required to form the 6-mer feature vector for each sample.
Figure \ref{fig:OldGrinderResults} compares the mean variational distance (VD) error 
at various taxonomic ranks as well as the algorithm execution time between 
SEK (${\rm OMP}_{\rm sek}^{+,1}$), Quikr and RDP's NBC.

As shown in figure \ref{fig:OldGrinderResults}, using the database $D_{\rm large}$, 
SEK outperforms both Quikr and RDP's NBC in terms of reconstruction error and has 
comparable execution time as Quikr. Both Quikr and SEK have significantly lower 
execution time than RDP's NBC. Using the database $D_{\rm small}$ (not shown here), 
SEK continues to outperform both Quikr and RDP's NBC in terms of reconstruction error, 
but only RDP's NBC in terms of execution time, as SEK had a median execution time of 
15.2 minutes versus Quikr's 25 seconds. All three methods have increasing error for 
lower taxonomic ranks, but the improvement of SEK over Quikr is emphasized for lower 
taxonomic ranks. 

\subsubsection{Results for second set of simulated data:}
\label{section:SimulatedDataResults_2}
\begin{figure}[!t]
	\centering
		(a){\includegraphics[width=\columnwidth]{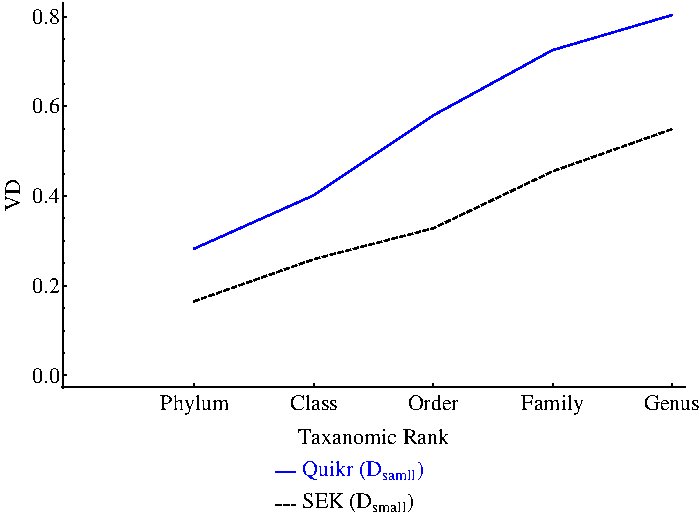}} 
		(b){\includegraphics[width=\columnwidth]{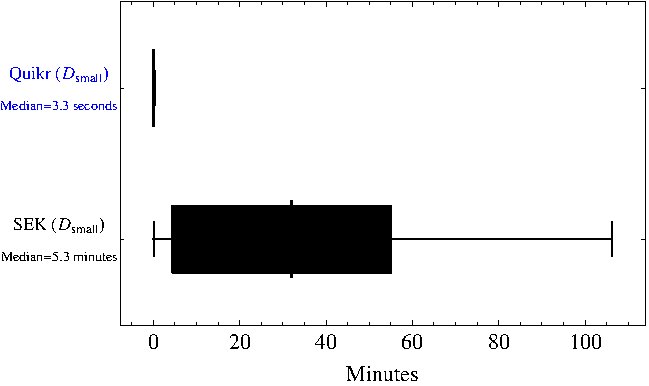}}
		\caption{For simulated data: Comparison of SEK (${\rm OMP}_{\rm sek}^{+,1}$) to Quikr on the 
		second set of simulated data. (a) Variational distance error averaged 
		over all 486 simulated datasets versus taxonomic rank for SEK and Quikr 
		trained using $D_{\rm small}$. (b) Algorithm execution time for SEK and 
		Quikr trained using $D_{\rm small}$. Whiskers denote range of the data, 
		vertical black bars designate the median, and the boxes demarcate quantiles.}
		\label{fig:VDNewGrinderExperiments}
\end{figure}

Figure \ref{fig:VDNewGrinderExperiments} summarizes the mean VD and algorithm 
execution time over the second set of simulated data described in section \ref{section:SimulatedData} 
for Quikr and SEK both trained on $D_{\rm small}$. 

Part (a) of Figure \ref{fig:VDNewGrinderExperiments} demonstrates that SEK shows much 
lower VD error in comparison to Quikr at every taxonomic rank. 
However, part (b) of Figure \ref{fig:VDNewGrinderExperiments} shows that this improvement 
comes at the expense of moderately increased mean execution time.

When focusing on the simulated datasets of length $100{\rm bp}$, $450{\rm bp}\pm 50{\rm bp}$, 
and $800{\rm bp}\pm 100{\rm bp}$, SEK had a mean VD of $0.803$, $0.410$, and $0.436$ respectively. 
As $L_w$ was set to $400$, this indicates the importance of choosing $L_w$ to roughly match 
the sequence length of a given sample when forming the $k$-mer training matrix if
sequence length is reasonably short (around 400 bp).

SEK somewhat experienced decreasing performance as a function of diversity: 
at the genus level, SEK gave a mean VD of 0.467, 0.579, and 0.603 for the 
simulated datasets with diversity 10, 100, and 500 respectively.

\subsection{Remarks on parameter choice and errors}
\label{subsec:Remarks_on_parameter_choice}
In SEK, we need to choose several parameters: $k$, $L_{w}$, $L_{p}$, $\nu$ and $I$.
Typically an increase in $k$ leads to better performance with the fact that a 
higher $k$ always subsumes a lower $k$ in the process of generating $k$-mers feature vectors.
The trend of improvement in performance with increase of $k$
was shown for Quikr \cite{Quikr_Koslicki_2013} and we believe that the same trend will also hold for SEK.
For SEK, the increase in $k$ results in exponential increase in row dimension of $\mathbf{X}$ matrix
and hence the complexity and memory requirement also increase exponentially. There is
no standard approach to fix $k$, except a brute force search. Let us now consider choice of $L_{w}$ and $L_{p}$. 
Our experimental results bring the following heuristic: choose $L_{w}$ to match the read length of sample data.
On the other hand, choose $L_{p}$ as small as possible to accommodate a high variability of
$k$-mers information in $\mathbf{X}$ matrix. A reduction in $L_{p}$ results to a linear
increase in column dimension of $\mathbf{X}$. Overall users should choose $k$, $L_{w}$ and $L_{p}$ such
that the dimension of $\mathbf{X}$ remains reasonable without considerable loss in estimation performance.
Finally we consider $\nu$ and $I$ parameters in Algorithm~\ref{alg:OMP_plus_1} that enforce sparsity,
with the aspect that computational complexity is $\mathcal{O}(4^{k} I^{3})$.
In general there is no standard automatic approach to choose these two parameters,
even for any standard algorithm. For example, the unconstrained Lagrangian form of LASSO mentioned in
section~\ref{subsec:Optimization_problem} also needs to set the parameter $\lambda$ by user.
For Algorithm~\ref{alg:OMP_plus_1}, $0< \nu <1$ should be chosen as a small positive number
and $I$ can be chosen as a fraction of row dimension of $\mathbf{X}$ that is $4^k$, of-course
with the requirement that $I$ is a positive integer. 
Let us choose $I=\lfloor \eta \times 4^k \rfloor$ where $0 < \eta \leq 1$. 
In case of a lower $k$, the system is more under-determined and naturally 
the enforcement of sparsity needs to be slackened to achieve a reasonable estimation performance.
Hence for a lower $k$, we need to choose a higher $\eta$ that can provide a good trade-off between complexity and estimation performance.
But, for a higher $k$, the system is less  under-determined and to keep the complexity reasonable, 
we should choose a lower $\eta$. Note that, for mock communities date, we used $k=4$ and $I=100$, and hence
$\eta = \frac{100}{4^4} \approx 0.4$, and for simulated data, we used $k=6$ and $I=409$, and hence 
$\eta = \frac{409}{4^6} \approx 0.1$. 

Further, it is interesting to ask what are the types of errors most common in SEK reconstruction.
In general, SEK reconstructs the most abundant taxa with remarkable fidelity. 
The less abundant taxa are typically more difficult to reconstruct and at times 
each behavior can be observed: low frequency taxa missing, miss-assigned, or their abundances miss-estimated.

\section{Discussion and Conclusion}

In this work we have shown that bacterial compositions of metagenomic samples can be
determined quickly and accurately from what initially appears to be very incomplete data.
Our method SEK uses only $k$-mer statistics of fixed length (here $k\sim 4,6$) of
reads from high-throughput sequencing data from the bacterial 16S rRNA genes to 
find which set of tens of bacteria are present out of a library of hundreds of species.
For a reasonable size of reference training data, the computational cost is dominated by 
the pre-computing of the $k$-mer statistics in the data and in the library; 
the computational cost of the central inference module is negligible, and can be 
performed in seconds/minutes on a standard laptop computer.

Our approach belongs to the general family of sparse signal processing where data sparsity
is exploited to solve under-determined systems. In metagenomics sparsity
is present on several levels. We have utilized the fact that 
$k$-mer statistics computed in windows of intermediate size 
vary substantially along the 16S rRNA sequences.
The number of variables representing the amount of reads assumed to be present in the
data from each genome and from each window is thus far greater than the number of
observations which are the $k$-mer statistics of all the reads in the data taken together.
More generally, while many bacterial communities are rich and diverse, the number
of species present in, for example the gut of one patient, will almost always be only a 
small fraction of the number of species present at the same position across a population, 
which in turn will only be a very small fraction of all known bacteria for which the
genomic sequences are available. We therefore believe that sparsity is a rather 
common feature of metagenomic data analysis which could have many applications
beyond the ones pursued here.

The major technical problem solved in the present paper stems from the fact that the
columns of the system matrix $\mathbf{X}$ linking feature vectors are highly correlated. 
This effect arises both from the construction of the
feature vectors i.e. that the windows are overlapping, and from biological similarity 
of DNA sequences along the 16S rRNA genes across a set of species.  
An additional technical complication is that the variables (species abundances) are
non-negative numbers and naturally normalized to unity, while in most methods 
of sparse signal processing there are no such constraints.
We were able to overcome these problems by constructing a new greedy algorithm based on orthogonal matching pursuit (OMP)
modified to handle the positivity constraint. The new algorithm, dubbed ${\rm OMP}_{\rm sek}^{+,1}$, integrates
ideas borrowed from kernel density estimators, mixture density models and sparsity-exploiting algebraic solutions.

During the manuscript preparation, we became aware that a similar methodology (Quikr)
has been developed by Koslicki et al in~\cite{Quikr_Koslicki_2013}. While there is a considerable similarity between Quikr and SEK, we note that Quikr is based only on sparsity-exploiting
algebraic solutions while SEK further exploits the additional sparsity assumption of 
non-uniform amplifications of variable regions in 16S rRNA sequences. 
Indeed, we hypothesize that the improvement of SEK over Quikr is mainly due to the superior 
training method of SEK.
The comparison between the two methods reported above in Figures~\ref{fig:SEK_comparison},
\ref{fig:OldGrinderResults} and \ref{fig:VDNewGrinderExperiments} shows that SEK performs generally better than Quikr.
The development of two new methodologies 
independently and roughly simultaneously reflect the timeliness and general interest of sparse 
processing techniques for bioinformatics applications.

\section*{Acknowledgement}

\paragraph{Funding\textcolon}
This work was supported by the Erasmus Mundus scholar program of the European Union (Y.L.), by the Academy of Finland
through its Finland Distinguished Professor program grant project 129024/Aurell (E.A.), ERC grant 239784 (J.C.) and 
the Academy of Finland Center of Excellence COIN (E.A. and J.C.),
by the Swedish Research Council Linnaeus Centre ACCESS (E.A., M.S., L.R., S.C. and M.V),
and by the Ohio Supercomputer Center and the 
Mathematical Biosciences Institute at The Ohio State University (D.K.).

\subsubsection{Conflict of interest statement.}

None declared. \newpage

\bibliographystyle{plain}
\bibliography{biblio_saikat_Pub,biblio_saikat_CS,biblio_saikat_Genomics}

\begin{thebibliography}{10}

\bibitem{cvx_toolbox_2013}
Cvx: A system for disciplined convex programming. http://cvxr.com/cvx/.
\newblock 2013.

\bibitem{COMPASS_Amir_2013}
A.~Amir, A.~Zeisel, O.~Zuk, M.~Elgart, S.~Stern, O.~Shamir, J.P. Turnbaugh,
  Y.~Soen, and N.~Shental.
\newblock {High-resolution microbial community reconstruction by integrating
  short reads from multiple 16S rRNA regions}.
\newblock {\em Nucleic Acids Res.}, 41(22):e205, 2013.

\bibitem{amir_CS_2011}
A.~Amir and O.~Zuk.
\newblock Bacterial community reconstruction using compressed sensing.
\newblock {\em J Comput Biol.}, 18(11):1723--41, 2011.

\bibitem{Angly2012}
F.~E. Angly, D.~Willner, F.~Rohwer, P.~Hugenholtz, and G.~W. Tyson.
\newblock Grinder: a versatile amplicon and shotgun sequence simulator.
\newblock {\em Nucleic acids research}, 40(12):e94, 2012.

\bibitem{Balzer2010}
S.~Balzer, K.~Malde, Lanz\'{e}n A., A.~Sharma, and I.~Jonassen.
\newblock Characteristics of 454 pyrosequencing data--enabling realistic
  simulation with flowsim.
\newblock {\em Bioinformatics}, 26(18):i420--5, 2010.

\bibitem{Bishop_MachineLearning_Book_2006}
C.~M. Bishop.
\newblock {\em Pattern recognition and machine learning}.
\newblock Springer, 2006.

\bibitem{Boyd_2004_Book}
S.~Boyd and L.~Vandenberghe.
\newblock {\em Convex optimization}.
\newblock Cambridge University Press, 2004.

\bibitem{ESPRIT_Tree_Cai_B16SrRNA_2011}
Y.~Cai and Y.~Sun.
\newblock Esprit-tree: hierarchical clustering analysis of millions of 16s rrna
  pyrosequences in quasilinear computational time.
\newblock {\em Nucleic Acids Research}, 39(14):e95, 2011.

\bibitem{CS_introduction_Candes_Wakin_2008}
E.J. Candes and M.B. Wakin.
\newblock An introduction to compressive sampling.
\newblock {\em IEEE Signal Proc. Magazine}, 25:21--30, march 2008.

\bibitem{Chatterjee_Sundman_Skoglund_2011_ICASSP_1}
S.~Chatterjee, D.~Sundman, and M.~Skoglund.
\newblock Look ahead orthogonal matching pursuit.
\newblock In {\em Acoustics, Speech and Signal Processing (ICASSP), 2011 IEEE
  International Conference on}, pages 4024 --4027, may 2011.

\bibitem{Chatterjee_Sundman_Vehkapera_Skoglund_TSP_2012}
S.~Chatterjee, D.~Sundman, M.~Vehkaper\"{a}, and M.~Skoglund.
\newblock Projection-based and look-ahead strategies for atom selection.
\newblock {\em Signal Processing, IEEE Transactions on}, 60(2):634 --647, feb.
  2012.

\bibitem{Cheng_Walker_Corander_2012metagenomic}
L.~Cheng, L.W. Walker, and J.~Corander.
\newblock Bayesian estimation of bacterial community composition from 454
  sequencing data.
\newblock {\em Nucleic Acids Research}, 2012.

\bibitem{Edgar_2010}
R.~C. Edgar.
\newblock Search and clustering orders of magnitude faster than blast.
\newblock {\em Bioinformatics}, 26(19):2460--2461, 2010.

\bibitem{Efron_2004_LARS}
B.~Effron, T.~Hastie, I.~Johnstone, and R.~Tibshirani.
\newblock Least angle regression.
\newblock {\em Ann. Statist.}, 32(2):407--499, 2004.

\bibitem{Haas_2011}
B.J. Haas, D.~Gevers, A.M. Earl, M.~Feldgarden, D.V. Ward, G.~Giannoukos,
  D.~Ciulla, D.~Tabbaa, S.K. Highlander, E.~Sodergren, B.~Methe, T.Z. DeSantis,
  Human~Microbiome Consortium, J.F. Petrosino, R.~Knight, and B.W. Birren.
\newblock Chimeric 16s rrna sequence formation and detection in sanger and
  454-pyrosequenced pcr amplicons.
\newblock {\em Genome Res.}, 21(3):494--504, 2011.

\bibitem{Makur_BAOMP_2011}
H.~Huang and A.~Makur.
\newblock Backtracking-based matching pursuit method for sparse signal
  reconstruction.
\newblock {\em Signal Processing Letters, IEEE}, 18(7):391--394, 2011.

\bibitem{Huson_MEGAN_2007}
D.H. Huson, A.F. Auch, J.~Qi, and S.C. Schuster.
\newblock Megan analysis of metagenomic data.
\newblock {\em Genome Res.}, 17(3):377--386, 2007.

\bibitem{Quikr_Koslicki_2013}
D.~Koslicki, S.~Foucart, and G.~Rosen.
\newblock Quikr: a method for rapid reconstruction of bacterial communities via
  compressive sensing.
\newblock {\em Bioinformatics}, 29(17):2096--2102, 2013.

\bibitem{meinicke2011mixture}
P.~Meinicke, K.P. A{\ss}hauer, and T.~Lingner.
\newblock Mixture models for analysis of the taxonomic composition of
  metagenomes.
\newblock {\em Bioinformatics}, 27(12):1618--1624, 2011.

\bibitem{Mitra_rRNA_MEGAN_2011}
S.~Mitra, M.~St\"ark, and D.H. Huson.
\newblock Analysis of 16s rrna environmental sequences using megan.
\newblock {\em BMC Genomics}, 2011.

\bibitem{Ong_2013}
S.H. Ong, V.U. Kukkillaya, A.~Wilm, C.~Lay, E.X.P. Ho, L.~Low, M.L. Hibberd,
  and N.~Nagarajan.
\newblock Species identification and profiling of complex microbial communities
  using shotgun illumina sequencing of 16s rrna amplicon sequences.
\newblock {\em PLoS One}, 8(4):e60811, 2013.

\bibitem{Stojnic_2010_Block_Sparse_JSTSP}
M.~Stojnic.
\newblock $l_2$/$l_1$-optimization in block-sparse compressed sensing and its
  strong thresholds.
\newblock {\em IEEE Journal of Selected Topics in Signal Processing},
  4(2):350--357, 2010.

\bibitem{Tropp_2007_OMP}
J.A. Tropp and A.C. Gilbert.
\newblock Signal recovery from random measurements via orthogonal matching
  pursuit.
\newblock {\em Information Theory, IEEE Transactions on}, 53(12):4655 --4666,
  dec. 2007.

\bibitem{von_Mering_2007}
C.~von Mering, P.~Hugenholtz, J.~Raes, S.~G. Tringe, T.~Doerks, L.~J. Jensen,
  N.~Ward, and P.~Bork.
\newblock Quantitative phylogenetic assessment of microbial communities in
  diverse environments.
\newblock {\em Science}, 315(5815):1126--1130, 2007.

\bibitem{NaiveBayesianClassifier_Wang_2007}
Q.~Wang, G.M. Garrity, J.M. Tiedje, and J.R Cole.
\newblock Na\"ive bayesian classifier for rapid assignment of rrna sequences
  into the new bacterial taxonomy.
\newblock {\em Appl. Environ. Microbiol.}, 73(16):5261--5267, 2007.

\bibitem{Wommack_read_lengths_2008}
K.E. Wommack, J.~Bhavsar, and J.~Ravel.
\newblock Metagenomics: read length matters.
\newblock {\em Appl Environ Microbiol.}, 74(5):1453--63, 2008.

\bibitem{Zuk_SPIRE_2013}
O.~Zuk, A.~Amir, A.~Zeisel, O.~Shamir, and N.~Shental.
\newblock {\em Accurate profiling of microbial communities from massively
  parallel sequencing using convex optimization}, volume LNCS 8214.
\newblock Springer, Cham, Switzerland, 2013.

\end{thebibliography}

\end{document}